\documentclass[pdflatex,sn-nature]{sn-jnl}

\usepackage{graphicx}%
\usepackage{multirow}%
\usepackage{amsmath,amssymb,amsfonts}%
\usepackage{amsthm}%
\usepackage{mathrsfs}%
\usepackage[title]{appendix}%
\usepackage{xcolor}%
\usepackage{textcomp}%
\usepackage{manyfoot}%
\usepackage{booktabs}%
\usepackage{algorithm}%
\usepackage{algorithmicx}%
\usepackage{algpseudocode}%
\usepackage{listings}%

\theoremstyle{thmstyleone}%

\theoremstyle{thmstyletwo}%

\theoremstyle{thmstylethree}%

\begin{document}

\title{Brillouin comb line amplifiers for optical atomic clocks}

\author*[1,2,3]{\fnm{Ryan L.} \sur{Russell}}\email{ryan.russell@sydney.edu.au}
\equalcont{These authors contributed equally to this work.}

\author[3,4]{\fnm{Lachlan} \sur{Pointon}}
\equalcont{These authors contributed equally to this work.}

\author*[3,4,5]{\fnm{Sarah Katherine} \sur{Scholten} \email{sarah.scholten@adelaide.edu.au}}
\author[1,2,3]{\fnm{Choon Kong} \sur{Lai}}
\author[1,2,3]{\fnm{Lisa-Sophie} \sur{Haerteis}}
\author[1,2]{\fnm{Cong Tinh} \sur{Bui}}
\author[1,2]{\fnm{Alvaro} \sur{Casas Bedoya}}
\author[3,4,5]{\fnm{Andreas} \sur{Boes}}
\author[6]{\fnm{Duk Yong} \sur{Choi}}
\author[6]{\fnm{Stephen J.} \sur{Madden}}
\author[3,4,5]{\fnm{Andre} \sur{Luiten}}
\author[1,2,3]{\fnm{Benjamin J.} \sur{Eggleton}}
\author*[1,2,3]{\fnm{Moritz} \sur{Merklein} \email{moritz.merklein@sydney.edu.au}}

\affil[1]{\orgname{Institute of Photonics and Optical Science (IPOS)}, 
\orgaddress{\city{Sydney}, \state{NSW}, \country{Australia}}}

\affil[2]{\orgname{The University of Sydney Nanoscience Institute (Sydney Nano)}, 
\orgaddress{\city{Sydney}, \state{NSW}, \country{Australia}}}

\affil[3]{\orgname{ARC Centre of Excellence in Optical Microcombs for Breakthrough Science (COMBS)}, 
\orgaddress{\country{Australia}}}

\affil[4]{\orgname{Institute for Photonics, Advanced Sensing and Quantum Technologies (IPAS-QT), Adelaide University}, 
\orgaddress{\city{Adelaide}, \state{SA}, \country{Australia}}}

\affil[5]{\orgname{Collaborative Research Hub for Alternative Positioning, Navigation and Timing, Adelaide University}, 
\orgaddress{\city{Adelaide}, \state{SA}, \country{Australia}}}

\affil[6]{\orgname{Laser Physics Centre, The Australian National University}, 
\orgaddress{\city{Canberra}, \state{ACT}, \country{Australia}}}

\abstract{

Optical atomic clocks outperform the microwave clocks at the heart of current timing and navigation infrastructure by orders-of-magnitude, and recent progress in on-chip integration provides the opportunity to widely deploy them. Conventionally, a narrow linewidth laser interrogates the atomic transition, and an optical frequency comb divides its stabilised frequency (hundreds of terahertz) down to a microwave that is electronically countable. The laser(s) impose a significant size, weight, power and cost penalty, which is a current barrier to low-cost, scalable, fully on-chip optical clocks. Here, we position the frequency comb as the central engine of the clock by using Brillouin-amplified comb lines for atomic spectroscopy, which eliminates the need for stable clock lasers.
Amplified comb lines reach an ultra-low frequency noise floor of $6\,\mathrm{Hz^2/Hz}$ at $30\,\mathrm{kHz}$ offset, an order-of-magnitude below a high-end commercial fibre laser. We realize a clock based on the $\text{5S}_{1/2}\rightarrow\text{5D}_{5/2}$ transition in $^{87}\mathrm{Rb}$ with competitive fractional frequency stability of $2 \times 10^{-13}$ at $\tau = 1$\,s, averaging to $9\times10^{-15}$ at $\tau=1000$\,s.
Lastly, we demonstrate a photonic chip-integrated Brillouin amplifier with a record 62.9\,dB gain and up to 40\,dB transferred noise suppression when pumped with a simple, low-cost diode laser. This opens a new route to compact optical clocks and enables use of atomic transitions lacking low-noise lasers by broadening a comb to the required wavelength and amplifying single lines.
}


\maketitle

\section*{Introduction}\label{sec1}
Optical atomic clocks (OACs) offer exquisite timing stability that can be leveraged for precision measurement, tests of fundamental physics and have the potential to inspire a revolution in modern navigation and timing technologies if made sufficiently compact and robust~\cite{ludlow2015,mcgrew2018,bothwell2022,degen2017}. OACs operate by transferring the frequency stability of an atomic transition down to countable radio frequencies. A spectroscopy laser is frequency stabilised to the atomic transition while an optical frequency comb (OFC) coherently divides its stable frequency from hundreds of terahertz, down to a countable radio frequency~\cite{diddams2010,fortier2019,newbury2011}. The laser is a complex component with footprint, frequency stability, tunability and reliability always in conflict, thus motivating a rethink of conventional OAC architectures.

Direct frequency-comb spectroscopy (DFCS) is an elegant approach that eliminates dedicated spectroscopy lasers altogether by dual-purposing the dedicated OFC as both the spectroscopic probe~\cite{martin2018,stern2020,marian2004,stowe2008,picque2019} and the optical-to-microwave divider. 
For example, the $\text{5S}_{1/2}\rightarrow\text{5D}_{5/2}$ transition in $^{87}\mathrm{Rb}$ used today in compact OACs~\cite{scholten2025,hilton2025demonstration} can be addressed by using two photons at 780\,nm and 776\,nm derived from a frequency-doubled OFC emitting in the near-infrared.
DFCS faces three key challenges distinct from conventional continuous-wave (CW) excitation that have so far limited its widespread use.
Firstly, for multi-photon transitions such as in $^{87}\mathrm{Rb}$, multiple comb-line pairs simultaneously satisfy the resonance condition~\cite{stowe2008}, producing a superposition of excitation pathways that complicates readout and introduces complex light shifts~\cite{felinto2009,stalnaker2012,hipke2014,dan2022,erickson2024,margolis2012,fendel2007two}.
Secondly, the low power per comb line (typically nanowatts) degrades the atomic fluorescence signal-to-noise ratio (SNR), and hence clock stability~\cite{newbury2011,picque2019}.
Finally, the finite temporal overlap of short OFC pulses within the atomic vapor reduces Doppler-cancellation efficiency, further degrading clock stability~\cite{stalnaker2012,hipke2014}. 
Existing approaches are not able to simultaneously address all three challenges without significant drawbacks.
For example, high-finesse Fabry–Perot filters can isolate individual comb lines but are passive, typically bulky, difficult to stabilize outside the laboratory, have a fixed periodic spectral response that makes alignment to the OFC difficult and require an additional external gain stage to amplify selected lines~\cite{gohle2007frequency,rutkowski2017continuous,erickson2024}. 
Microcombs relax the required filter resolution due to their larger comb line spacing, but sacrifice fine control over detuning from intermediate states that exist within atomic transitions (e.g. $\text{5P}_{3/2}$ in $^{87}\mathrm{Rb}$), typically lack cavity-length actuation, have low power-per-line and are limited by thermorefractive noise~\cite{stowe2008,kippenberg2018,drake2020thermal}.

\begin{figure}[t]
    \centering
    \includegraphics[width=1\linewidth]{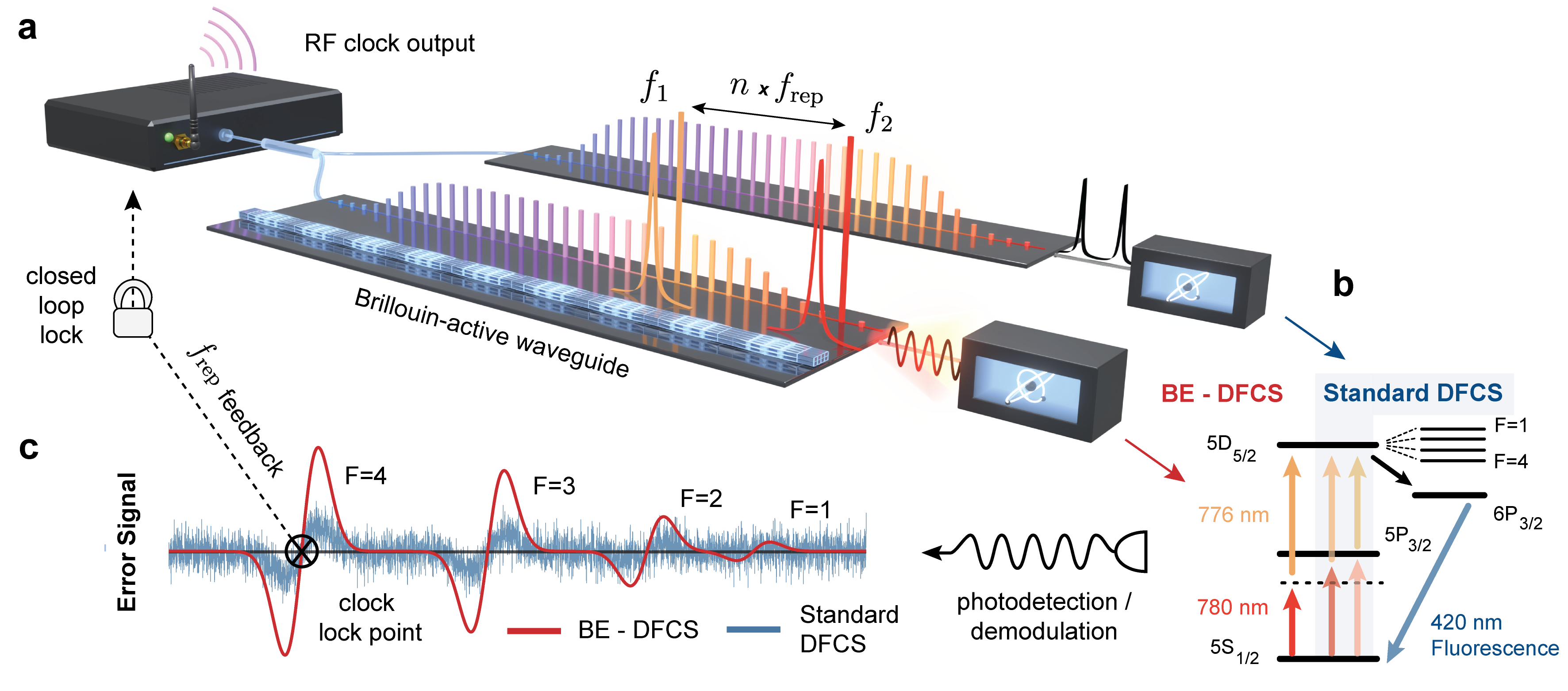}
    \caption{\textbf{Optical atomic clock based on Brillouin-enhanced direct frequency comb spectroscopy (BE-DFCS).} \textbf{a)} With BE-DFCS, two continuous-wave SBS amplified comb lines at $f_1$ and $f_2$, separated by $n\times f_{\mathrm{rep}}$, drive the two-photon $\text{5S}_{1/2}\rightarrow\text{5D}_{5/2}$ transition of rubidium-87 ($^{87}\mathrm{Rb}$). In contrast, standard DFCS involves pulsed-excitation of the atoms, with all combinations of weak comb lines driving the transition (2 combinations illustrated here). 
    \textbf{b)} $^{87}\mathrm{Rb}$ energy levels are shown, including $\text{F}=1$ to $\text{F}=4$ hyperfine transition levels. Excitation by comb lines at $f_1$ and $f_2$ induce 420\,nm fluorescence with four hyperfine features corresponding to  $\text{F}=1 - 4$. 
    \textbf{c)} Fluorescence is demodulated to yield clock error signals for BE-DFCS (red, high signal-to-noise) compared to standard DFCS (blue, poor signal-to-noise). Actuation of $f_{\mathrm{rep}}$ steers $f_1$ and $f_2$ onto the transition to stabilize the clock. $f_{\mathrm{rep}}$ is photodetected and can be used as a stable RF clock output.}
    \label{fig:fig1}
\end{figure}

Amplifying individual comb lines with stimulated Brillouin scattering (SBS) in a single-pass waveguide addresses all three challenges facing DFCS, while remaining simple to incorporate into a range of photonic systems, including OACs.
The Brillouin gain bandwidth is ultra-narrow --- on the order of 40\,MHz --- which is narrower than typical OFC repetition rates (50\,-\,200\,MHz), allowing a single comb line to be selectively amplified to milliwatt level~\cite{rohde2013,terra2019,galindo2016,xiong2024,xiao2025fiber,yao2024precise}. 
The amplifier transfers energy to existing comb lines in a cavity-free configuration, in contrast to a conventional laser that generates new optical carriers and must be independently stabilized \cite{loh2020operation}.
Brillouin amplifiers have been integrated on photonic chips to support system SWaP reduction, with many SBS-capable platforms now emerging, e.g. chalcogenide, silica, silicon nitride and lithium niobate~\cite{eggleton2019brillouin,eggleton2011chalcogenide,merklein2022100}. 
However, all current on-chip amplifiers provide orders-of-magnitude weaker net gain than atomic spectroscopy with an OFC requires.

Here, we demonstrate Brillouin-enhanced DFCS (BE-DFCS) and integrate it into a two-color rubidium-87 ($^{87}\mathrm{Rb}$) clock with a fractional frequency stability of $2 \times 10^{-13}$ at integration time $\tau = 1$\,s, averaging down to $\sim 9\times10^{-15}$ at $\tau=1000$\,s, competitive with the state-of-the-art for compact OACs.
The two SBS-amplified comb lines driving the $\text{5S}_{1/2}\rightarrow\text{5D}_{5/2}$ transition of $^{87}\mathrm{Rb}$, exhibit a low frequency-noise floor of $6\,\mathrm{Hz}^2/\mathrm{Hz}$ at 30\,kHz offset, an order of magnitude below that of a commercial ultra-stable fibre laser and in agreement with a quadrature-resolved noise theory.
In a leap towards system integration, an integrated on-chip amplifier is demonstrated with a record Brillouin gain exceeding 60\,dB for $^{87}\mathrm{Rb}$ spectroscopy. Pump frequency noise transfer through the on-chip SBS amplifier is suppressed by up to 40\,dB, allowing a simple, low-cost distributed-feedback (DFB) laser to capably replace high-quality spectroscopy lasers.
As OFCs and Brillouin gain are largely wavelength-agnostic, any photonic system incorporating an OFC could swap high-quality lasers for an SBS comb amplifier driven by compact, low-cost diode lasers. In all, our architecture offers a promising route to OACs with low size, weight, power and cost (SWaP-C), and can be extended to atomic transitions unsupported by commercial low-noise lasers.

\begin{figure}[b]
    \centering
    \includegraphics[width=1\linewidth]{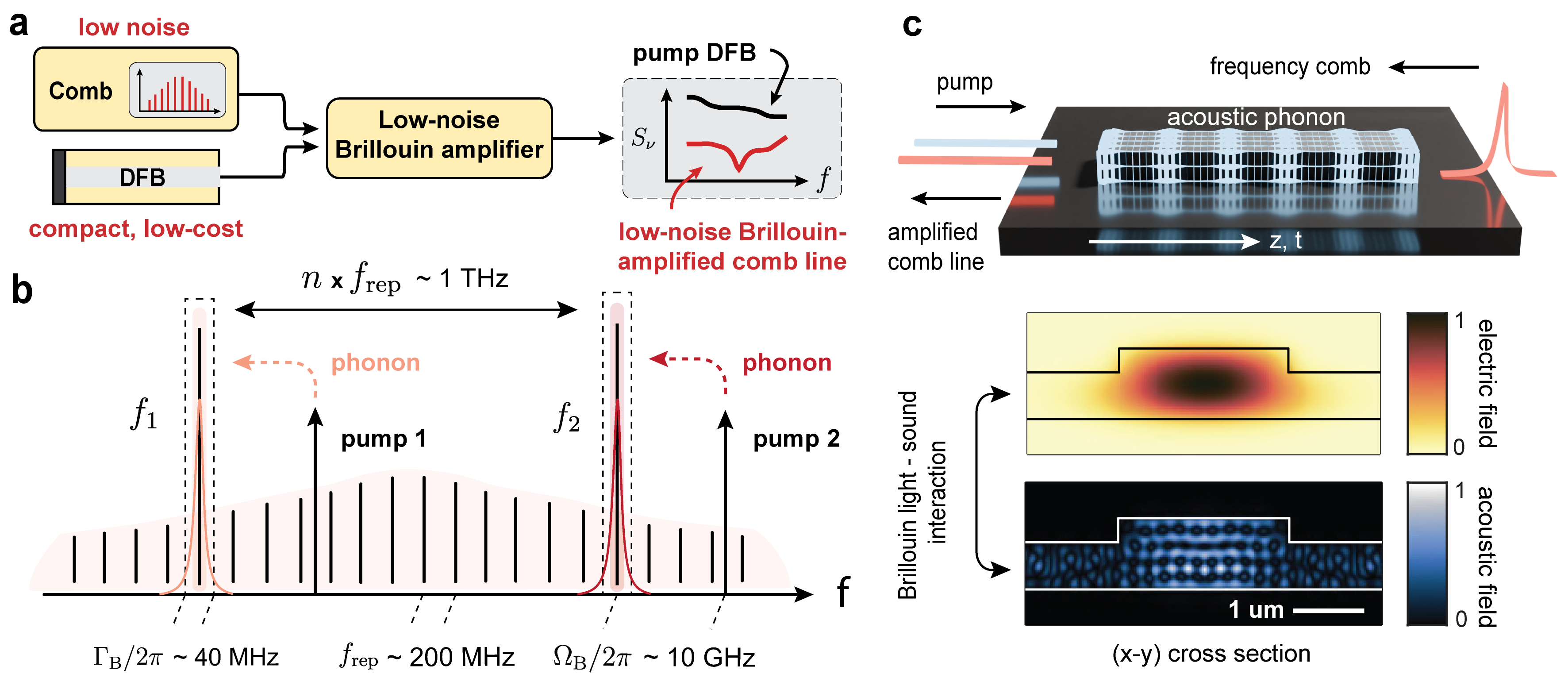}
    \caption{\textbf{Principle of low-noise frequency comb amplification with a Brillouin amplifier.} \textbf{a)} Concept of a Brillouin amplifier preserving low-noise OFC input while pumped by a low-cost DFB laser with high noise. Inset: phase noise of the DFB pump (black) and the amplified comb line (red) -- $S_\nu$ is phase noise power spectral density.
    \textbf{b)} Spectral representation of SBS amplification of comb lines $f_1$ and $f_2$, illustrating the order-of-magnitude differences between the Brillouin frequency shift $\Omega_{\mathrm{B}}/2\pi\sim10$\,GHz, gain bandwidth $\Gamma_{\mathrm{B}}/2\pi\sim40$\,MHz and the comb-line spacing $f_{\mathrm{rep}}\sim200$\,MHz.
    \textbf{c)} Spatiotemporal $(z,t)$ SBS interaction between the pump, the counter-propagating comb and the acoustic phonon in a waveguide. Inset: simulated $(x\text{-}y)$ cross-sections of the interacting (i) electromagnetic and (ii) acoustic modes. Scale bar, 1\,$\mu$m.}
    \label{fig:fig2}
\end{figure}

\section*{Results}\label{sec2}

\subsection*{Brillouin-enhanced direct comb spectroscopy for atomic clocks}

In our BE-DFCS clock architecture, two Brillouin-amplified comb lines at $f_1$ and $f_2$, separated by $\Delta_f=n\times f_{\mathrm{rep}}$ (\autoref{fig:fig1}\,a) are frequency-doubled and drive the two-photon $\text{5S}_{1/2}\rightarrow\text{5D}_{5/2}$ transition in $^{87}\mathrm{Rb}$ (780/776\,nm). This occurs via a single, well-defined excitation pathway rather than the many simultaneously-resonant comb-line pairs of standard DFCS (\autoref{fig:fig1}\,b), yielding a clean, hyperfine-resolved fluorescence spectrum and a low-noise clock error signal (\autoref{fig:fig1}\,c). 
Since the amplified comb line separation ($\Delta_f$) is strictly a multiple of $f_{\mathrm{rep}}$, the clock can be fully stabilized by actuating the fast piezoelectric stretcher within the OFC oscillator.
\autoref{fig:fig2}\,a conceptually shows a Brillouin amplifier seeded by a low-noise comb line and driven by a low-quality DFB pump laser with high phase noise. 
In this configuration, the system outputs CW light inheriting the excellent phase coherence of the seed comb, rather than the noisy pump. 
In the frequency domain (\autoref{fig:fig2}\,b), each pump is blue-detuned from its target line $f_{1,2}$ by the Brillouin frequency shift of the gain medium ($\Omega_{\mathrm{B}}/2\pi\sim10$\,GHz), with gain mediated by a $\Gamma_{\mathrm{B}}/2\pi\sim40$\,MHz phonon resonance.
Since this is an order of magnitude narrower than a typical OFC repetition rate ($f_{\mathrm{rep}}\sim200$\,MHz), a single targeted line is amplified while its neighbours remain outside the Brillouin gain band, and are unamplified.
In the spatiotemporal picture (\autoref{fig:fig2}\,c), counter-propagating pump and comb electromagnetic fields couple to each other via guided acoustic phonons which are transversely confined in the waveguide cross-section.

\begin{figure}[htbp]
    \centering
    \includegraphics[width=1\linewidth]{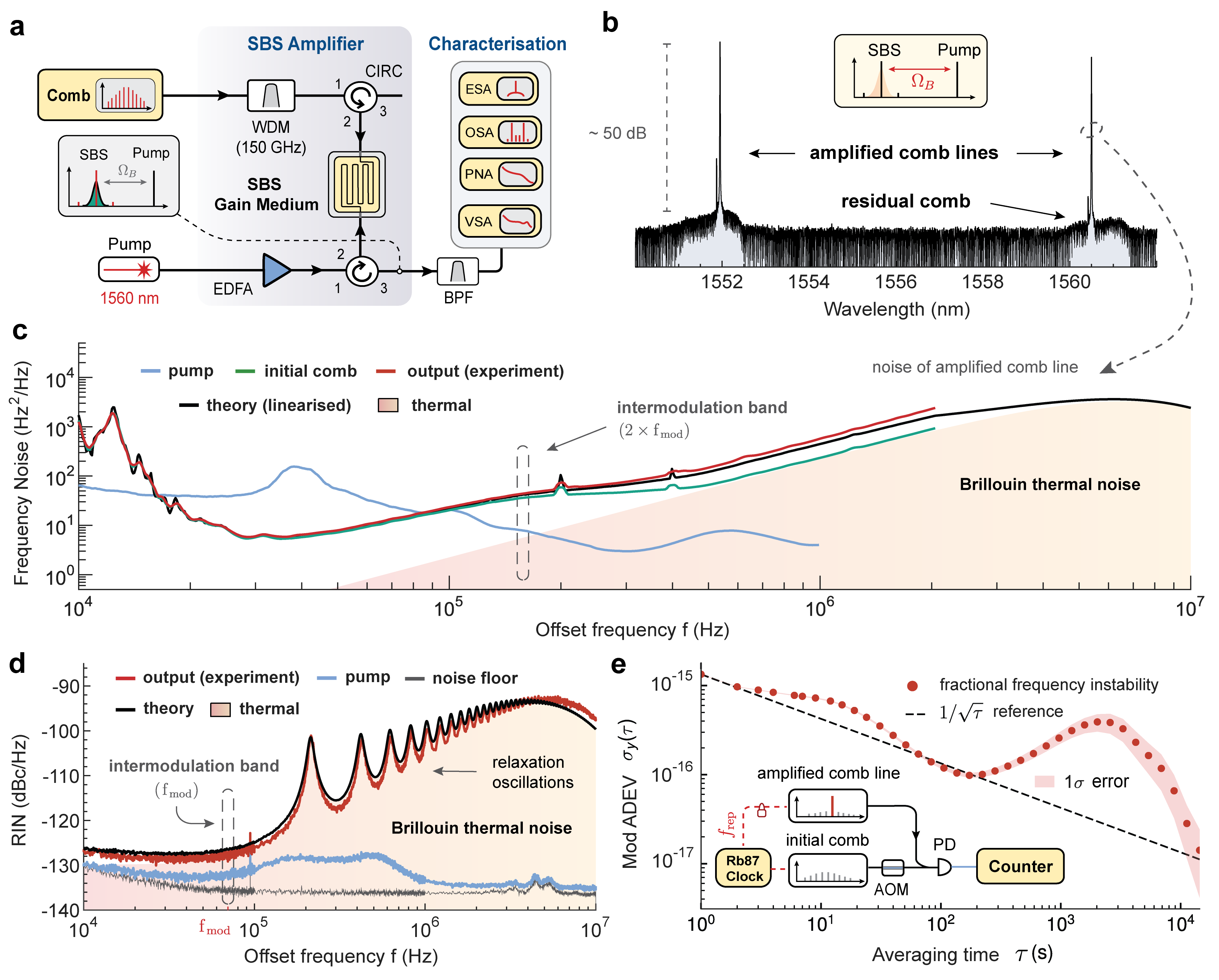}
    \caption{\textbf{Brillouin comb-line amplification in highly nonlinear fibre.}
\textbf{a)} Experimental setup for SBS comb-line amplification and characterization. 
\textbf{b)} Optical spectrum of the amplifier output, with amplified comb lines at 1552\,nm and 1560\,nm and the residual comb. Inset: SBS gain and pump, separated by the Brillouin frequency shift $\Omega_{\mathrm{B}}$.
\textbf{c)} Frequency noise of the pump (blue), initial comb (green) and SBS output (red), with the linearized theory (black) and Brillouin thermal noise (shaded). 
\textbf{d)} RIN of the SBS output (red), theory (black), pump (blue) and noise floor (grey), with the thermal limit (shaded).
\textbf{e)} Modified Allan deviation of the amplified comb line (red points; shaded, $1\sigma$ error) with a $1/\sqrt{\tau}$ reference (dashed). Inset: measurement setup. 
WDM: wavelength division multiplexer, CIRC: circulator, EDFA: erbium doped fibre amplifier, BPF: bandpass filter, ESA\,/\,OSA: electrical/optical spectrum analyser, PNA: phase noise analyser, VSA: vector signal analyser, AOM: acousto-optic modulator, PD: photodetector.}
    \label{fig:fig3}
\end{figure}

\subsection*{Low-noise Brillouin comb-line amplifier}

We first demonstrate low-noise Brillouin amplification of individual comb lines in a 510\,m long highly nonlinear fibre (HNLF), before integrating the amplifier on-chip. 
As shown in the experimental setup (\autoref{fig:fig3}\,a), the OFC is counterpropagated against the pump and a single comb line is amplified.
This setup is simply extended to the amplification of two comb lines by introducing two pumps, which will be used for the two-color $^{87}\mathrm{Rb}$ clock demonstration. 
The resulting output is displayed as an optical spectrum in \autoref{fig:fig3}\,b, showing $\sim$\,50\,dB SBS gain, with the residual unamplified comb remaining close to the instrument noise floor. 

\autoref{fig:fig3}\,c shows the frequency-noise (FN) performance of the SBS comb-line amplifier.
The OFC used is a compact, self-referenced, passively mode-locked laser based on the NIST design \cite{sinclair2015invited}, with repetition rate stabilized to a rubidium OAC \cite{scholten2025}.
FN of the Brillouin amplified comb line closely reproduces that of the input comb line (green), remaining strongly decoupled from the input pump noise (blue), which sits approximately an order-of-magnitude higher between $20\,\text{kHz} < f < 100\,\text{kHz}$.
Numerical simulations of the Brillouin stochastic differential equations provide insight into the underlying noise mechanisms (\textcolor{blue}{Supplementary Materials}) by resolving total SBS amplifier noise into amplitude- and phase-quadrature components from each noise source. 
Simulations corroborate the finding that pump FN transfer to the amplified comb line is negligible. 
Instead, it is thermally-seeded spontaneous Brillouin scattering that establishes a white FN limit (shaded orange region in \autoref{fig:fig3}\,c), extending to the Brillouin gain bandwidth ($\Gamma_{\mathrm{B}}/2\pi\sim45$\,MHz).
We formalize this transition to the thermally-limited regime by defining a crossover frequency, $f_{\mathrm{crit}}$, at which the thermal-noise floor intersects the input probe's noise spectrum,
\begin{equation}
    f_{\mathrm{crit}}
  \;=\;\sqrt{\frac{\Delta\nu\,P_s^{\rm in}}{\pi(n_{\mathrm{{th}}}+1)\hbar\omega_s}} \sim 420\;\mathrm{kHz},
\end{equation}
where $\Delta\nu \sim 132$\,Hz is the intrinsic Lorentzian linewidth of the probe, $P_s^{\rm in} \sim 350$\,nW is the input probe power, $n_{\mathrm{th}} \sim 643$ is the thermal occupancy of the 9.71\,GHz phonon mode at 300\,K, $\hbar$ is the reduced Planck constant, and $\omega_s \sim 2\pi \times 192\,$THz is the optical angular frequency.
$f_{\mathrm{crit}}$ is equivalently the offset frequency beyond which the input probe's technical noise sits within the amplifier's noise figure of the shot-noise limit. 
As will be discussed in the next section, atomic clock stabilization involves imparting a modulation on the spectroscopy light at frequency $f_{\mathrm{mod}} \sim 80$\,kHz. To prevent intermodulation effects from degrading clock stability, we must achieve minimal phase noise degradation of the amplified comb line(s) used for spectroscopy at $2\times f_{\mathrm{mod}} \sim 160$\,kHz. 
Since $2\times f_{\mathrm{mod}} \ll  f_{\mathrm{crit}}$ in our system, Brillouin thermal noise negligibly contributes to clock instability.

Also key to our approach is that the relative intensity noise (RIN) remains low at $f_{\mathrm{mod}}$, achieved through Brillouin gain self-regulation (\autoref{fig:fig3}\,d, red curve).
Stokes intensity fluctuations locally alter pump depletion, and this pump perturbation counterpropagates to adjust the SBS gain against the original fluctuation after a round-trip delay of $2L/v$, where $v$ is the group velocity. 
Slow (low-frequency) fluctuations are therefore suppressed, strongly attenuating the amplified thermal noise contribution below $\sim100$\,kHz.
Faster fluctuations outpace the delayed correction and manifest as relaxation oscillations (ROs) at harmonics of the inverse round-trip time $f_{\mathrm{RO}} = v/2L \sim 200$\,kHz, in agreement with simulation (\autoref{fig:fig3}\,d, black curve).
The 510\,m HNLF was chosen such that $f_{\mathrm{mod}} < f_{\mathrm{RO}}$ and consequently the enhanced RIN does not strongly couple to clock instability.

Long-term frequency stability of the amplified comb-line can be degraded by time-varying strain and temperature gradients over the long fibre~\cite{terra2019}. 
In our case, the fractional instability of the amplifier is markedly low ($1\times10^{-15}$ at $\tau = 1$\,s), averaging down to $4\times10^{-17}$ at $\tau = 10^4$\,s (modified Allan deviation normalized to the 192\,THz optical carrier, \autoref{fig:fig3}\,e).
If required for a specific application, we expect that long-term stability can be improved using shielded or actively stabilized optical fibre platforms and by on-chip device integration.

\subsection*{Optical atomic clock demonstration}

We now demonstrate an optical frequency standard based on the BE-DFCS architecture, using two-color (780\,nm and 776\,nm) excitation of the $\text{5S}_{1/2}\rightarrow\text{5D}_{5/2}$ two-photon transition of $^{87}$Rb (\autoref{fig:fig1}). 
Since the two-photon transition rate ($W$) scales quadratically with detuning from the intermediate $\text{5P}_{3/2}$ state ($\Delta_{\mathrm{int}}$) as $W\propto 1/\Delta_{\mathrm{int}}^2$, independent control of the two optical frequencies provides a key advantage of two-color excitation over the conventional single-color (778\,nm) scheme.
A nominal detuning nearly three orders of magnitude smaller ($\Delta_{\mathrm{int}} \sim$\,1.5\,GHz) for the two-color case compared to the fixed $\Delta_{\mathrm{int}} \sim$\,1.1\,THz of the single-color case, means a given excitation rate can be achieved with far lower optical power and vapor cell temperature. 
This relaxes the demand for optical amplification, while mitigating light- and pressure-induced frequency shifts that compromise long-term clock stability.

\begin{figure}[htbp]
    \centering
    \includegraphics[width=1\linewidth]{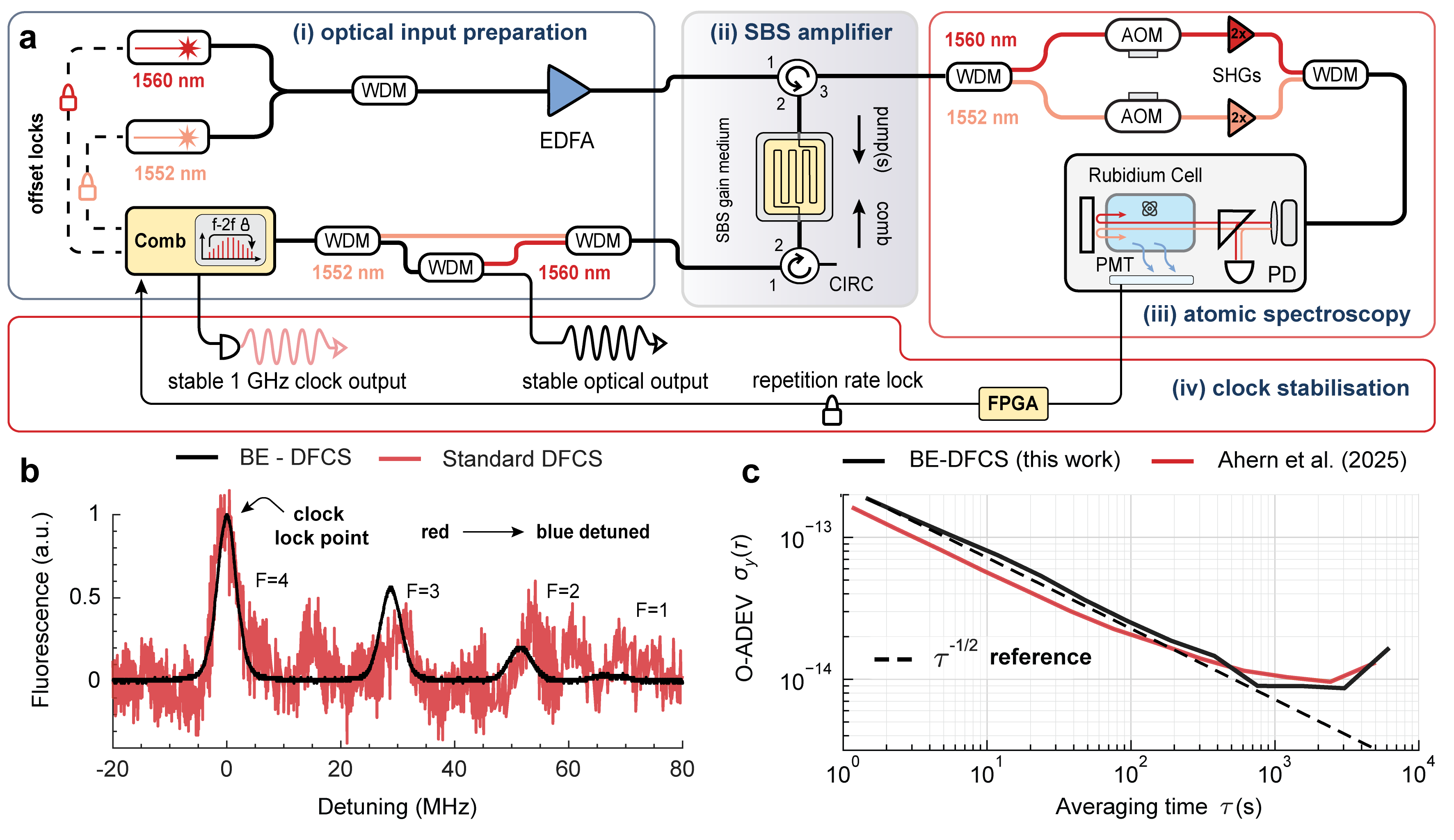}
    \caption{\textbf{Two-color rubidium optical atomic clock based on BE-DFCS.} \textbf{a)} Experimental setup, comprising (i) optical inputs: two fibre lasers at 1552\,/\,1560\,nm and an OFC, (ii) SBS amplifier, (iii) spectroscopy and (iv) clock stabilization, with stable optical and 1\,GHz RF clock outputs. \textbf{b)} Fluorescence versus detuning from the clock transition for BE-DFCS (black) and standard DFCS (red), showing the excitation of $\text{F}=1$ to $\text{F}=4$ hyperfine levels of $\text{5D}_{5/2}$. Amplitudes are normalized. \textbf{c)} Fractional frequency instability of the BE-DFCS clock (black) compared to a state-of-the-art reference two-color Rb clock (red)~\cite{scholten2025}, compared to a $1 / \sqrt{\tau}$ ADEV reference line (dashed). FPGA: field programmable gate array. }
    \label{fig:fig4}
\end{figure}

\autoref{fig:fig4}\,a shows the setup. 
Two OFC lines at 1560\,nm and 1552\,nm are amplified via SBS as described above, separated by a wavelength-division multiplexer (WDM), individually power-controlled and modulated by AOMs, frequency-doubled to 780\,nm and 776\,nm respectively, then co-injected into a rubidium vapor cell. 
Fluorescence at 420\,nm from the $\text{6P}_{3/2}\rightarrow\text{5S}_{1/2}$ decay path is demodulated to yield an error signal, and a servo actuates the fast piezo inside the OFC oscillator to steer $f_\mathrm{rep}$ and thus stabilize both amplified comb lines to the two-photon resonance.
Due to the milliwatt-level power in each amplified comb line, the fluorescence SNR is substantially higher than that of standard DFCS (\autoref{fig:fig4}\,b).
The clock output is taken either optically (any comb tooth) near 1556\,nm or as a 1\,GHz RF signal from the fifth harmonic of $f_\mathrm{rep}$. Detailed operating conditions and parameters are in \textcolor{blue}{Methods}.

The overlapping Allan deviation (ADEV) of the clock (\autoref{fig:fig4}\,c) is $1.9\times10^{-13}$ at 1.5\,s and averages down as $1 / \sqrt{\tau}$ for approximately 1000\,s, reaching $8.6\times10^{-15}$ at $\tau = 3000$\,s. 
This is competitive with leading single-color~\cite{martin2018,Newman2021} and two-color~\cite{scholten2025} two-photon rubidium clocks, the latter shown in \autoref{fig:fig4}\,c after reprocessing with an overlapping ADEV.
In a frequency-modulated lock, detection shot noise, laser FN and laser RIN can each contribute a white $1 / \sqrt{\tau}$ ADEV floor, and we evaluate each contribution separately (\textcolor{blue}{Supplementary Materials}).
Laser FN at $2f_\mathrm{mod} = 160$\,kHz couples to clock instability through the intermodulation effect~\cite{Audoin1991}, while laser RIN at $f_\mathrm{mod}$ affects the error-signal gradient, and so manifests as an apparent frequency noise.
Using the measured SBS-output FN of $S_\nu = 44$\,Hz$^2$/Hz at $2\times f_{\mathrm{mod}}$ and RIN of $-127$\,dBc/Hz at $f_{\mathrm{mod}}$(\autoref{fig:fig3}\,c,\,d, red curves), the fractional ADEV contributions are $1.7\times10^{-14}\,\tau^{-1/2}$ from FN and $7.7\times10^{-15}\,\tau^{-1/2}$ from RIN, at least an order of magnitude below the measured clock stability $\sim 2\times10^{-13}\,\tau^{-1/2}$.
The clock is therefore shot-noise limited, and SBS amplifier noise is not itself a limiting factor of clock performance.

\subsection*{On-chip integration}
\begin{figure}[htbp]
    \centering
    \includegraphics[width=1\linewidth]{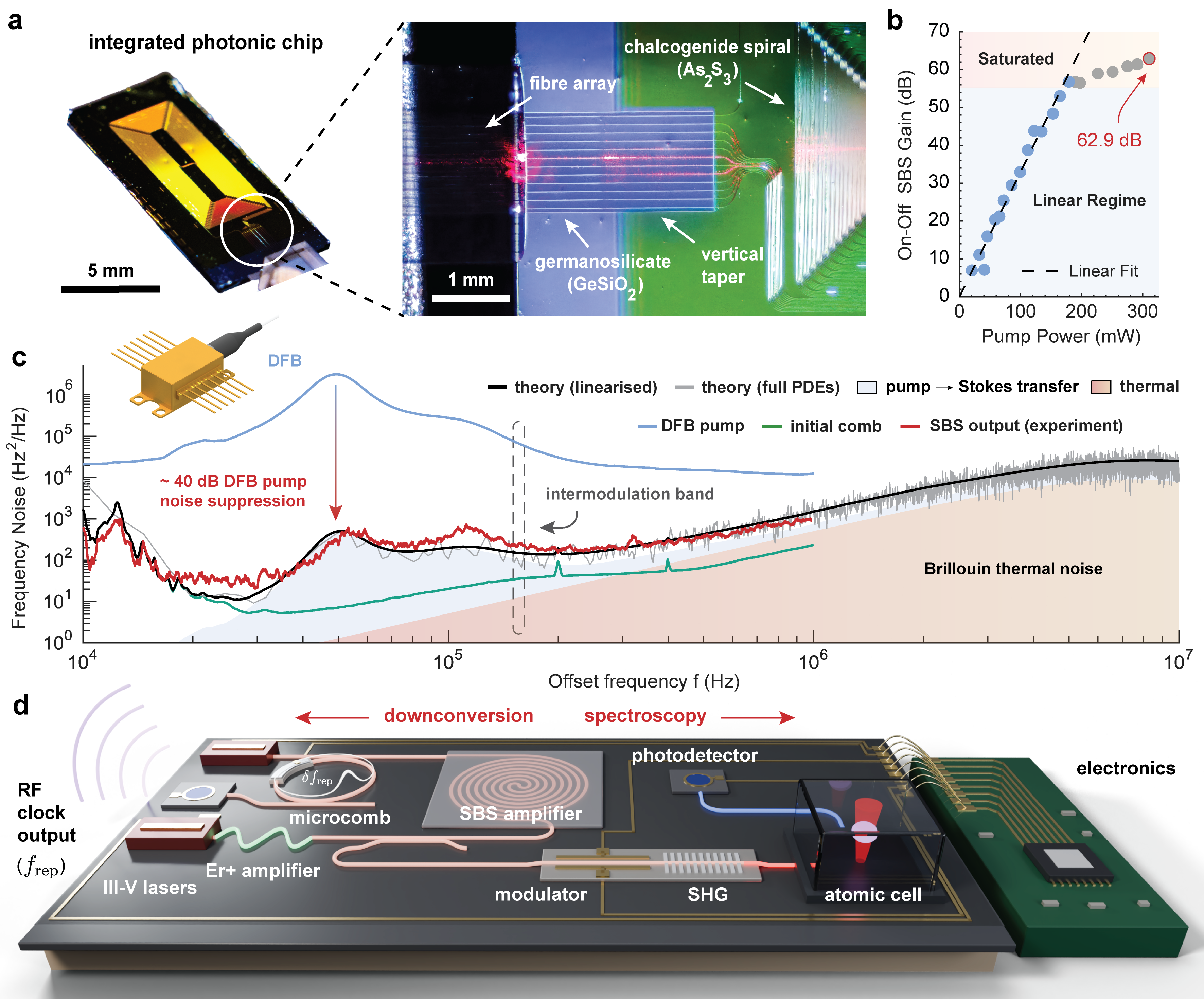}
    \caption{\textbf{On-chip integrated Brillouin comb-line amplifier.}  
\textbf{a)} Photograph of the integrated photonic chip (scale bar, 5\,mm). Inset: zoomed-in optical microscope image (scale bar, 1\,mm) showing the fibre array bonded to mode-matched GeSiO$_2$ waveguides ($n\sim1.44$) before vertically tapering into the SBS-active As$_2$S$_3$ waveguides ($n\sim 2.43$) that are packaged in a compact spiral containing six 20\,cm long waveguides within a $0.8$cm$^2$ total footprint.
\textbf{b)} On-off SBS gain versus on-chip pump power in the linear and saturated regimes (blue and grey points, respectively), with a linear fit (dashed).
\textbf{c)} Frequency noise of an amplified comb line with a simple low-cost DFB laser as pump for the SBS amplifier: DFB pump (blue), initial comb (green), SBS output (red), linearized theory (black) and full stochastic PDE simulation (grey). Shaded regions: pump-to-Stokes transfer (blue) and thermal contributions (orange).
\textbf{d)} Concept of a fully chip-integrated optical clock based on Brillouin-enhanced DFCS. Feedback on the repetition rate ($\delta f_{\text{rep}}$) of a microcomb source enables closed-loop stabilisation of individual Brillouin-amplified comb lines to the atomic reference. Optical-to-microwave clock downconversion is achieved by photodetection of the stabilised microcomb $f_{\text{rep}}$. 
}
    \label{fig:fig5}
\end{figure}

In this section, we take a major step towards a chip-scale BE-DFCS clock by demonstrating that the SBS amplifier component can be implemented on a photonic chip with an 0.8\,cm$^2$ footprint (\autoref{fig:fig5}\,a) and maintain its low-noise performance when pumped by a simple high-noise DFB laser.
A 20\,cm As$_2$S$_3$ waveguide with 7.6\,dB propagation loss and high SBS gain coefficient of $G_{\mathrm{B}}\sim804$\,m$^{-1}$W$^{-1}$ is hybrid integrated with germanosilicate (GeSiO$_2$) waveguides~\cite{lai2022hybrid} for low-loss and robust interfacing with the bonded fibre array (inset), achieving 10.9\,dB total device insertion loss.
We report a peak on/off gain of 62.9\,dB (\autoref{fig:fig5}\,b).
The corresponding net gain far exceeds standing records in integrated Brillouin photonics (\autoref{tab:sbs_platforms}). Accounting for propagation loss, it is 15.3\,dB higher than standard As$_2$S$_3$ waveguides~\cite{choudhary2017giant}, 36.8\,dB higher than hybrid As$_2$S$_3$-on-Si waveguides~\cite{morrison2017hybrid} and significantly higher than all current state-of-the-art silicon (Si)~\cite{kittlaus2016silicon}, silicon nitride (Si$_3$N$_4$)~\cite{klaver2026saw}, and lithium niobate (LiNbO$_3$)~\cite{yu2025tfln} devices.
Our compact on-chip device provides sufficient net SBS gain to directly amplify nW-level comb lines for $^{87}\mathrm{Rb}$ spectroscopy (\textcolor{blue}{Supplementary Materials}).

\begin{table}[htbp]
\centering
\footnotesize
\setlength{\tabcolsep}{4pt}
\caption{State of the art for integrated (on-chip) Brillouin amplifiers across various waveguide and material platforms. On/off gain is the signal gain with pump on versus off. Net$_\alpha$ and Net$_\mathrm{IL}$ are the net Brillouin gains considering waveguide propagation loss alone or the total insertion loss, respectively. n/r denotes a figure not reported in the referenced work.}
\label{tab:sbs_platforms}
\begin{tabular}{@{}lccccc@{}}
\toprule
Platform & $L$ (cm) & On/off gain (dB) & Net$_\alpha$ (dB) & Net$_\mathrm{IL}$ (dB) & $G_{\mathrm{B}}$ (m$^{-1}$W$^{-1}$) \\
\midrule
\textbf{As$_2$S$_3$-GeSiO$_2$ (this work)} & \textbf{20.0} & \textbf{62.9} & \textbf{55.3} & \textbf{52.0} & \textbf{804.0} \\
As$_2$S$_3$~\cite{choudhary2017giant} & 23.0 & 52.0 & 40.0 & 32.0 & n/r \\
As$_2$S$_3$-on-Si, hybrid~\cite{morrison2017hybrid} & 5.8 & 22.5 & 18.5 & n/r & 750.0 \\
Si, suspended membrane~\cite{kittlaus2016silicon} & 2.9 & 6.9 & 5.2 & n/r & 1120.0 \\
Si$_3$N$_4$, SAW-enhanced~\cite{klaver2026saw} & 6.7 & 10.0 & 5.0 & n/r & 81.0 \\
LiNbO$_3$, suspended (TFLN)~\cite{yu2025tfln} & 3.0 & 8.5 & 6.4 & n/r & 129.5 \\
\midrule
HNLF (fibre reference, this work) & 51000 & 62.4 & 61.9 & 60.7 & 0.8 \\
\bottomrule
\end{tabular}
\end{table}

Beyond the self-evident SWaP advantage, on-chip integration of SBS devices is further motivated by the fact that environmental perturbations such as thermal gradients and strain are imprinted as phase noise on the amplified comb-line output, that scales with interaction length \cite{Terra2010, xiong2025low, terra2019}.
Hence, the reduced interaction lengths associated with on-chip integration minimize these adverse environmental effects. We show that the added frequency noise is as low as $10^{-3}$\,Hz$^2$/Hz at 10\,Hz offset and remains below 1\,Hz$^2$/Hz up to a $\sim$\,10\,kHz offset where mechanical and acoustic noise is most dominant (\textcolor{blue}{Methods, Supplementary Materials}). 
Another distinct advantage of the on-chip amplifier is that the output RIN can be reduced close to the pump level ($\sim -125$\,dBc/Hz) when operated in deep saturation (\textcolor{blue}{Supplementary Materials}).
The significant in-band RIN penalty associated with low-frequency relaxation oscillations, as seen in fibre-based amplifiers, is avoided completely (\autoref{fig:fig3}\,d) owing to the 1000\,x shorter interaction length.

Lastly, we characterize the on-chip amplifier pumped with a low-cost DFB laser with high phase noise to motivate future laser-on-chip integration.
Transfer of this DFB pump noise (\autoref{fig:fig5}\,c, blue curve) to the amplified comb line (red curve) is suppressed by up to 40\,dB, again in excellent agreement with simulation. 
The measured spectrum is reproduced both by the linearized transfer-function model (black curve, \textcolor{blue}{Supplementary Materials}) and by direct time-domain integration of the full stochastic PDEs (grey curve, \textcolor{blue}{Methods}).
Hence, we suggest that SBS comb-line amplifiers pumped by low-cost diodes could replace high-quality laser sources in a range of metrology systems that employ frequency combs, of which OACs are a primary candidate.

\newpage

\section*{Discussion}\label{sec12}

Our BE-DFCS architecture repositions the frequency comb as the central engine of the entire optical clock system, rather than merely a frequency divider.
Amplifying single comb lines via stimulated Brillouin scattering permits the comb to perform two roles which conventional clocks delegate into separate subsystems: atomic spectroscopy and optical-to-microwave downconversion.
Brillouin amplification strongly preserves the comb's phase coherence, allowing high SWaP-C ultra-stable lasers to be replaced with Brillouin-active waveguides pumped by cheap diode lasers.
Clock stabilization is also much simpler since the amplified comb lines probing the transition are inherently pinned to the comb’s repetition rate, which is used as the RF clock output.

The on-chip amplifier demonstrated in this work establishes a realistic vision for a fully chip-integrated BE-DFCS clock (visualized in \autoref{fig:fig5}\,d), with the required material platforms now rapidly maturing. 
Ultra-low loss silicon nitride (Si$_3$N$_4$) is a preferred CMOS-compatible base layer in which soliton microcombs are most mature~\cite{kippenberg2018, liu2020monolithic}. 
Self-referencing, piezo-actuation and thermorefractive noise are key challenges, but such microcombs could ultimately be the engines of on-chip BE-DFCS clocks. High-gain Brillouin amplification can be achieved by depositing chalcogenide glasses directly onto the Si$_3$N$_4$, similar to the recent demonstration with tellurium oxide~\cite{klaver2026saw}. 
Thin-film lithium niobate (TFLN) can be bonded at the wafer-scale \cite{churaev2023heterogeneously} or be micro-transfer printed~\cite{vanackere2023heterogeneous}, providing the $\chi^{(2)}$ nonlinearity required for high-efficiency second-harmonic generation~\cite{wang2018ultrahigh} that converts amplified near-infrared comb lines into the visible.
Acousto-optic~\cite{wan2022highly} or electro-optic~\cite{wang2018integrated} modulators can then be co-integrated, mirroring our recent progress on chalcogenide-TFLN Brillouin circuits for microwave photonics applications~\cite{lai2026brillouin}.
Since we have shown that Brillouin amplifiers tolerate high pump phase noise, simple III-V on-chip lasers can suffice~\cite{xiang20233d} and erbium-implantation of the Si$_3$N$_4$ waveguides enables pump amplification~\cite{liu2022photonic}. 
Finally, microfabricated rubidium vapor cells bring the atomic spectroscopy to chip scale~\cite{hummon2018photonic,newman2019architecture,Newman2021}. Hence, this heterogeneous architecture contains all key elements of an all-on-chip BE-DFCS clock: comb generation, conventional wideband gain, narrowband Brillouin gain, frequency conversion, modulation and atomic spectroscopy.

In all, we expect the simplicity and performance of our BE-DFCS architecture to pave the way for a new generation of low SWaP-C optical clocks suited to operational environments such as aerospace and maritime.

\newpage 

\section*{Methods}\label{methods}

\subsection*{SBS amplifier characterization}

The Brillouin gain spectrum of each amplifier (510\,m HNLF from Optical Fiber Solutions (OFS) and a 20\,cm on-chip waveguide) was measured using a pump-probe technique (\textcolor{blue}{Supplementary Materials}).
A single CW laser (Teraxion PureSpectrum-LM) was split into pump and probe arms.
In the pump arm, the upper sideband generated by a Mach-Zehnder intensity modulator (MZM; Fujitsu H74M-5208) was selected by a narrow bandpass filter (Alnair BVF-300CL), amplified by a pre-amplifier and booster EDFA in cascade (Amonics AEDFA-PA-35-B-FA and AEDFA-33-B-FA) then coarsely filtered (Santec OTF-320) to remove broadband amplified spontaneous emission. In the probe arm, a dual-parallel MZM (DP-MZM; JDSU 540161B-Q) driven by the swept RF output of a vector network analyser (Agilent N5224A) generated a single upper sideband to act as the SBS probe, which is attenuated to $-34$\,dBm ($\sim 400$\,nW) at the input of the device under test (DUT).
The carrier-to-sideband ratio in this arm is $\sim\,11$\,dB. Pump and probe were counter-propagated through the DUT via circulators. 
The SBS-amplified probe and the residual transmitted laser carrier were heterodyned on a photodiode (Optilab PD-28-C-DC) connected to the VNA input via an RF amplifier (Minicircuits ZVA-213-S+). 
The measured S$_{21}$ response traces the Brillouin gain spectrum as the pump-probe detuning is swept across a frequency range significantly larger than the gain linewidth.

\subsection*{Phase noise measurements}

Optical phase noise was characterized by one of two methods depending on the source. For CW sources, namely the pump lasers and SBS amplified comb line output, a delayed self-heterodyne (DSH) interferometry technique was used. The phase noise of a single comb line was instead determined by direct heterodyne detection against a lower-noise reference laser. 
The amplified comb line noise (\autoref{fig:fig3}\,c, red) was also cross-checked with this method. 
In both cases, the phase noise analyser (PNA) records the single-sideband phase noise $\mathcal{L}(f)$ of an RF beat note, which is converted to the one-sided phase noise and frequency noise power spectral densities, $S_\phi(f) = 2\times10^{\mathcal{L}(f)/10}$ and $S_\nu(f) = f^2 S_\phi(f)$, respectively.

Using DSH, the optical signal was split between the two arms of an unbalanced Mach-Zehnder interferometer.
One arm contained $\sim$\,100\,m of polarization-maintaining fibre, imposing a delay of $\tau = 495$\,ns ($1/\tau = 2.02$\,MHz).
This fibre delay was thermally and vibrationally isolated to suppress environmental noise that limits the noise floor at low offset frequencies. The other arm is frequency shifted by an acousto-optic modulator ($f_\mathrm{AOM} = 83$\,MHz).
Recombination with the delayed arm produced an RF beat whose phase noise was measured with a digital PNA (Koheron ALPHA250). 
The beat note phase is the difference between the laser phase at times separated by the delay, $\Delta\phi(t) = \phi(t) - \phi(t-\tau)$ and hence the measured noise power spectral density is simply related to that of the laser phase noise $S_\phi(f)$ via the interferometer transfer function
\begin{equation}
    S_{\Delta\phi}(f) = 4\sin^2(\pi f\tau)\,S_\phi(f),
    \label{eq:dshi}
\end{equation}

Alternatively using the direct heterodyne method, the optical test signal e.g. the comb was heterodyned against an ultra-stable reference laser, with beat detected on a photodiode (Thorlabs PDB415C).
The phase noise of the beat was measured with a PNA (Microchip 53100A, or PN2060D), and since the optical fields are uncorrelated, it can be expressed as the sum of two contributions,
\begin{equation}
    S_{\phi,\mathrm{beat}}(f) = S_{\phi}(f) + S_{\phi,\mathrm{ref}}(f),
    \label{eq:het}
\end{equation}
Hence, this measurement approximates the true signal phase noise wherever the reference is substantially quieter, and otherwise provides an upper bound. 
The low-noise reference is a stimulated Brillouin laser with a white frequency-noise floor of $\sim 0.1$\,Hz$^2$/Hz, stabilized to an ultra-low expansion cavity (ULE) (\textcolor{blue}{Supplementary Materials}).

\subsection*{RIN measurements}

Relative intensity noise was measured by direct detection on a photodiode (Thorlabs PDB415C) operated in unbalanced mode with $\sim$\,30\,$\mu$W optical power to avoid saturation. 
The detector output was split equally between a multimeter measuring the DC voltage level $V_\mathrm{DC}$ and a vector signal analyser (Agilent 89410A) measuring the voltage noise spectral density $V(f)$.
The optical RIN was then obtained by normalizing to the DC level,
\begin{equation}
\mathrm{RIN}(f) = 10\log_{10}\!\left[\left(V(f)/V_\mathrm{DC}\right)^2\right]
\end{equation}

\subsection*{SBS amplifier long-term frequency stability}

The long-term frequency instability contributed by the HNLF amplifier (\autoref{fig:fig3}\,e) was measured by splitting the OFC into two arms.
In one arm, a single comb line was amplified by the SBS amplifier and in the other, the unamplified comb was frequency-shifted by an acousto-optic modulator (AOM) driven at $f_{\mathrm{AOM}}\sim40$\,MHz by an RF source (Keysight 33500B) referenced to a stable cryogenic sapphire oscillator (CSO). 
The two arms were recombined on a balanced photodiode (Thorlabs PDB415C), and the beat note at $f_{\mathrm{AOM}}$ was counted directly against the AOM drive signal (Microchip 53100A). 
Since the comb is common to both arms, the counted frequency fluctuations $\delta f(t)$ capture the instability of the SBS amplifier together with the non-common fibre paths of the interferometer.
Fractional instability is referenced to the optical carrier $\nu_{\mathrm{opt}}\sim192$\,THz, and is reported as the modified Allan deviation.

The CSO instability contribution scales as $f_{\mathrm{AOM}}/\nu_{\mathrm{opt}}$ which imposes a stability floor of $\sigma_{y,\mathrm{floor}}\sim\sigma_{\mathrm{CSO}}\,(f_{\mathrm{AOM}}/\nu_{\mathrm{opt}})\sim10^{-21}$ for a nominal CSO fractional instability of $\sigma_{\mathrm{CSO}}\sim3\times10^{-15}$. Therefore, the CSO contribution is not limiting.

\subsection*{On-chip SBS amplifier added noise}

The frequency noise added by the 20\,cm on-chip SBS amplifier was quantified by splitting the comb into two arms, similar to previous studies of conventional semiconductor amplifiers~\cite{teyssieux2026phase}: 1) a reference arm, and 2) a test arm containing the SBS amplifier and an AOM driven at $f_{\mathrm{RF}}$ (setup in \textcolor{blue}{Supplementary Materials}).
Environmental noise imprints along the gain medium in the test arm only, and balanced detection of the two arms isolates this non-common-mode contribution.

\subsection*{Rubidium optical atomic clock}

The clock was operated with $P_{776} = 2.20$\,mW and $P_{780} = 100\,\mu$W (at the atoms), an intermediate-state detuning of $\Delta_\mathrm{int} = 1.5$\,GHz, and a cell temperature of 70\,$^\circ$C. The 1552\,nm and 1560\,nm pump lasers were frequency stabilized to a fixed frequency offset via respective phase-locked loops to their nearest OFC line using a balanced photodiode (Thorlabs PDB415C) and an FPGA servo acting on the laser piezos. 
The 1552\,nm light was frequency-modulated at $f_\mathrm{mod} = 80$\,kHz using an AOM. 
The detected and demodulated fluorescence from the $\textrm{6P}_{3/2}\rightarrow\text{5S}_{1/2}$ $^{87}$Rb decay path provided the error signal for an FPGA servo that steered $f_\mathrm{rep}$ via the piezoelectric actuators in the OFC oscillator. 
The OFC's carrier-envelope offset frequency was stabilized by $f$--$2f$ self-referencing. 
Optical power was monitored by a photodiode (Thorlabs FDS100) immediately before the atoms and stabilized by FPGA feedback to the two AOMs.
Since both colors overlap on this photodiode (PD), they were distinguished by time-gating in order to provide independent feedback. 
The 780\,nm beam was square-wave amplitude-modulated at $f_\mathrm{AC} \ll f_\mathrm{mod}$ and its demodulated amplitude stabilized, while the 776\,nm beam power loop maintains the DC setpoint level on the PD. 
The effects of residual amplitude modulation (RAM) at $f_\mathrm{mod}$ were reduced by detecting its in-phase and quadrature components on the same PD and applying a tone of equal amplitude and opposite phase to the AOM RF drive amplitude~\cite{scholten2025}.

The fluorescence spectra in \autoref{fig:fig4}\,b were obtained by ramping $f_\mathrm{rep}$, which sweeps the sum frequency of the offset-locked comb lines across the two-photon resonance. 
Similarly, the standard DFCS trace used a broad $f_\mathrm{rep}$ scan, of which one of the repeating fluorescence features is shown.

Clock stability was measured from the beat note between the 1560\,nm OFC line and an NKT Basik E15 laser stabilized to an ultra-low-expansion (ULE) cavity, detected on a balanced photodiode (Thorlabs PDB415C).
The beat frequency was recorded with an Agilent 53132A counter.

\subsection*{Numerical simulation model}

Brillouin amplifier noise was modelled using stochastic coupled-mode equations describing the spatiotemporal evolution of the pump $a_p(z,t)$, counter-propagating Stokes (comb line) $a_s(z,t)$, and the acoustic field $b(z,t)$,
\begin{align}
  \frac{\partial b}{\partial t} + \left(\frac{\Gamma_{\mathrm{B}}}{2}-\mathrm{i}\Delta\right) b
    &= \mathrm{i}g\,a_p a_s^{*} + \xi(z,t), \label{eq:m-b}\\
  \frac{\partial a_p}{\partial z} + \frac{1}{v}\frac{\partial a_p}{\partial t}
    &= -\frac{\alpha}{2}a_p + \mathrm{i}g\,a_s b, \label{eq:m-ap}\\
  -\frac{\partial a_s}{\partial z} + \frac{1}{v}\frac{\partial a_s}{\partial t}
    &= -\frac{\alpha}{2}a_s + \mathrm{i}g\,a_p b^{*}, \label{eq:m-as}
\end{align}
where $|a_{p,s}|^2$ are the optical field powers, $v$ is the optical group velocity, $\alpha$ the propagation loss, $\Gamma_{\mathrm{B}}$ the phonon decay rate and $\Delta$ the detuning from the Brillouin resonance.
Acoustic transport is neglected, since the phonon decay length is much shorter than the device length. $\xi$ is the thermal Langevin force driving the phonon mode.

Equations~\eqref{eq:m-b}--\eqref{eq:m-as} can be integrated directly in the time domain using the measured pump and probe noise spectra as boundary inputs; this yields the full output response of the system and accurately replicates experimental results. 
To distinguish the noise contributions by their independent physical origins e.g. thermal versus pump transfer noise, the stochastic PDEs were linearized about the steady-state solutions, yielding explicit transfer functions for both noise quadratures (phase\,/\,intensity). 
The output phase noise can be conveniently expressed as the incoherent sum
\begin{equation}
  S_{\phi,s}^{\mathrm{out}}(f) = \left|H_{p\to s}\right|^2 S_{\phi,p}^{\mathrm{in}}(f) + \left|H_{s\to s}\right|^2 S_{\phi,s}^{\mathrm{in}}(f) + S_{\phi,\mathrm{th}}(f),
  \label{eq:m-total}
\end{equation}
with an analogous expression for intensity.
The thermal contribution $S_{\phi,\mathrm{th}}$ is obtained by propagating $\xi$ through the system. The full derivation is provided in \textcolor{blue}{Supplementary Materials}.

\subsection*{SBS amplifier chip fabrication}

The As$_2$S$_3$-GeSiO$_2$ chips were fabricated at wafer scale using a heterogeneous integration process flow.
A 1\,$\mu$m As$_2$S$_3$ film was thermally evaporated onto a GeSiO$_2$ layer through a shadow mask, forming $\sim$\,1.3\,mm vertical tapers for adiabatic mode transfer between the GeSiO$_2$ and As$_2$S$_3$ waveguides. 
After coating with a 150\,nm SU-8 protective layer, the film was thermally annealed (130\,$^\circ$C, 24\,h) and then annealed with green light for 24 hours.
20\,cm long As$_2$S$_3$ rib waveguides, compactified into a spiral geometry, were patterned by UV lithography and partially etched 330\,nm via inductively coupled plasma reactive ion etching (ICP-RIE) using CHF$_3$. 
A 30\,nm TiO$_2$ layer was then deposited by atomic layer deposition. A second lithography step defined 3\,$\mu$m-wide GeSiO$_2$ waveguides beneath the 2.2\,$\mu$m-wide As$_2$S$_3$ waveguides, followed by a multi-step reactive ion etch through the TiO$_2$, As$_2$S$_3$ and GeSiO$_2$ layers. The wafer was clad with a 10\,$\mu$m thick low-index polymer (ZPU, Chemoptics; $n=1.44$ at 1550\,nm) before cleaving into individual photonic chips. A polarization maintaining fibre array was subsequently bonded to the chip using a UV-cured resin.

\section*{Acknowledgements}\label{acknowledgements}

We thank the Adelaide University portable optical atomic clocks team and past and current members of the Precision Measurement Group at Adelaide University within the Institute for Photonics, Advanced Sensing and Quantum Technologies for their assistance. We thank the Optofab (SA), NSW and ACT nodes of the Australian National Fabrication Facility (ANFF), and in particular E. Johnson, A. Dowler and L. Thomas, for technical support. R.L.R thanks Prof. Michael Steel (Macquarie University, Australia) for insightful discussions. Some parts of this work regarding the two-color rubidium frequency standard are covered by US patent 10353270 B2.

\section*{Declarations}

\subsection*{Funding} 

This work was sponsored by the US Office of Naval Research (project number N00014-24-1-2009) and the Australian Research Council Centre of Excellence in Optical Microcombs for Breakthrough Science (project number CE230100006) supported by the Australian Government.
The Commonwealth of Australia (represented by the Advanced Strategic Capabilities Accelerator) supports this research through a Defence Science Partnerships agreement.
This project and publication is supported in part by the Australian Government Department of Education through Australia's Economic Accelerator Program.
S.K.S. is the recipient of an Australian Research Council Australian Early Career Researcher Industry Fellowship (project number IE240100056) funded by the Australian Government. 
R.L.R. acknowledges the support of an Australian Government Research Training Program (RTP) Scholarship. 

The views, opinions and findings expressed are those of the authors and should not be interpreted as representing the official views or policies of the Department of Defence or the US Government.

\subsection*{Conflict of interests}

A.L. is a founder and the managing director of QuantX Labs, and maintains equity interests. All other authors declare no competing interests.

\subsection*{Data availability}

The data that support the findings of this study are available from the corresponding author(s) upon reasonable request.

\subsection*{Author contribution}
R.L.R., S.K.S. and M.M. conceived the work.
R.L.R. and L.P. designed and conducted the experiments, with assistance from S.K.S., L.-S.H. and M.M.
R.L.R., C.K.L., C.T.B., A.C.B., D.Y.C., S.J.M. and M.M. contributed to the design, fabrication and/or characterization of the on-chip amplifier.
L.P. and S.K.S. contributed to the design and implementation of the clock architecture.
R.L.R. and M.M. conducted the theoretical and numerical analysis of Brillouin noise.
R.L.R., L.P., S.K.S. and M.M. contributed to the analysis and interpretation of the data.
R.L.R. and L.P. wrote the manuscript with input from all authors.
S.K.S., A.B., A.L., B.J.E. and M.M. contributed to supervision and funding acquisition.

\bibliography{sn-bibliography}%

\backmatter

\bmhead{Supplementary information}
This manuscript is accompanied by a separate Supplementary Information file.

\end{document}